\ifx\newheadisloaded\relax\immediate\write16{***already loaded}\endinput\else\let\newheadisloaded=\relax\fi
\gdef\isonnarrowscreen{F}
\gdef\PSfonts{T}
\magnification\magstep1

\newdimen\papwidth
\newdimen\papheight
\newskip\beforesectionskipamount  
\newskip\sectionskipamount 
\def\sectionskip{\vskip\sectionskipamount}
\def\beforesectionskip{\vskip\beforesectionskipamount}
\papwidth=16truecm
\if F\isonnarrowscreen
\papheight=22truecm
\voffset=0.4truecm
\hoffset=0.4truecm
\else
\papheight=22truecm
\voffset=-1.5truecm
\hoffset=-1truecm
\fi
\hsize=\papwidth
\vsize=\papheight
\catcode`\@=11
\ifx\amstexloaded@\relax
\else
\nopagenumbers
\headline={\ifnum\pageno>1 {\hss\tenrm-\ \folio\ -\hss} \else
{\hfill}\fi}
\fi
\catcode`\@=\active
\newdimen\texpscorrection
\texpscorrection=0.15truecm 

\def\sectionsize{\twelvepoint}
\def\sectiontype{\bf}
\def\subsectionsize{}
\def\subsectiontype{\bf}
\def\em{\sl}
\newfam\truecmsy
\newfam\truecmr
\newfam\msbfam
\newfam\scriptfam
\newfam\frakfam
\newfam\frakbfam

\newskip\ttglue 
\if T\isonnarrowscreen
\papheight=11.5truecm
\fi
\if F\PSfonts
\font\twelverm=cmr12
\font\tenrm=cmr10
\font\eightrm=cmr8
\font\sevenrm=cmr7
\font\sixrm=cmr6
\font\fiverm=cmr5

\font\twelvebf=cmbx12
\font\tenbf=cmbx10
\font\eightbf=cmbx8
\font\sevenbf=cmbx7
\font\sixbf=cmbx6
\font\fivebf=cmbx5

\font\twelveit=cmti12
\font\tenit=cmti10
\font\eightit=cmti8
\font\sevenit=cmti7
\font\sixit=cmti6
\font\fiveit=cmti5

\font\twelvesl=cmsl12
\font\tensl=cmsl10
\font\eightsl=cmsl8
\font\sevensl=cmsl7
\font\sixsl=cmsl6
\font\fivesl=cmsl5

\font\twelvei=cmmi12
\font\teni=cmmi10
\font\eighti=cmmi8
\font\seveni=cmmi7
\font\sixi=cmmi6
\font\fivei=cmmi5

\font\twelvesy=cmsy10	at	12pt
\font\tensy=cmsy10
\font\eightsy=cmsy8
\font\sevensy=cmsy7
\font\sixsy=cmsy6
\font\fivesy=cmsy5
\font\twelvetruecmsy=cmsy10	at	12pt
\font\tentruecmsy=cmsy10
\font\eighttruecmsy=cmsy8
\font\seventruecmsy=cmsy7
\font\sixtruecmsy=cmsy6
\font\fivetruecmsy=cmsy5

\font\twelvetruecmr=cmr12
\font\tentruecmr=cmr10
\font\eighttruecmr=cmr8
\font\seventruecmr=cmr7
\font\sixtruecmr=cmr6
\font\fivetruecmr=cmr5

\font\twelvebf=cmbx12
\font\tenbf=cmbx10
\font\eightbf=cmbx8
\font\sevenbf=cmbx7
\font\sixbf=cmbx6
\font\fivebf=cmbx5

\font\twelvett=cmtt12
\font\tentt=cmtt10
\font\eighttt=cmtt8

\font\twelveex=cmex10	at	12pt
\font\tenex=cmex10

\font\twelvemsb=msbm10	at	12pt
\font\tenmsb=msbm10
\font\eightmsb=msbm8
\font\sevenmsb=msbm7
\font\sixmsb=msbm6
\font\fivemsb=msbm5


\font\tenfrm=eufm10
\font\eightfrm=eufm8
\font\sevenfrm=eufm7
\font\sixfrm=eufm6
\font\fivefrm=eufm5

\font\tenfrb=eufb10
\font\eightfrb=eufb8
\font\sevenfrb=eufb7
\font\sixfrb=eufb6
\font\fivefrb=eufb5
\font\twelvescr=eusm10 at 12pt
\font\tenscr=eusm10
\font\eightscr=eusm8
\font\sevenscr=eusm7
\font\sixscr=eusm6
\font\fivescr=eusm5
\fi
\if T\PSfonts
\font\twelverm=ptmr	at	12pt
\font\tenrm=ptmr	at	10pt
\font\eightrm=ptmr	at	8pt
\font\sevenrm=ptmr	at	7pt
\font\sixrm=ptmr	at	6pt
\font\fiverm=ptmr	at	5pt

\font\twelvebf=ptmb	at	12pt
\font\tenbf=ptmb	at	10pt
\font\eightbf=ptmb	at	8pt
\font\sevenbf=ptmb	at	7pt
\font\sixbf=ptmb	at	6pt
\font\fivebf=ptmb	at	5pt

\font\twelveit=ptmri	at	12pt
\font\tenit=ptmri	at	10pt
\font\eightit=ptmri	at	8pt
\font\sevenit=ptmri	at	7pt
\font\sixit=ptmri	at	6pt
\font\fiveit=ptmri	at	5pt

\font\twelvesl=ptmro	at	12pt
\font\tensl=ptmro	at	10pt
\font\eightsl=ptmro	at	8pt
\font\sevensl=ptmro	at	7pt
\font\sixsl=ptmro	at	6pt
\font\fivesl=ptmro	at	5pt

\font\twelvei=cmmi12
\font\teni=cmmi10
\font\eighti=cmmi8
\font\seveni=cmmi7
\font\sixi=cmmi6
\font\fivei=cmmi5

\font\twelvesy=cmsy10	at	12pt
\font\tensy=cmsy10
\font\eightsy=cmsy8
\font\sevensy=cmsy7
\font\sixsy=cmsy6
\font\fivesy=cmsy5
\font\twelvetruecmsy=cmsy10	at	12pt
\font\tentruecmsy=cmsy10
\font\eighttruecmsy=cmsy8
\font\seventruecmsy=cmsy7
\font\sixtruecmsy=cmsy6
\font\fivetruecmsy=cmsy5

\font\twelvetruecmr=cmr12
\font\tentruecmr=cmr10
\font\eighttruecmr=cmr8
\font\seventruecmr=cmr7
\font\sixtruecmr=cmr6
\font\fivetruecmr=cmr5


\font\twelvett=cmtt12
\font\tentt=cmtt10
\font\eighttt=cmtt8

\font\twelveex=cmex10	at	12pt
\font\tenex=cmex10

\font\twelvemsb=msbm10	at	12pt
\font\tenmsb=msbm10
\font\eightmsb=msbm8
\font\sevenmsb=msbm7
\font\sixmsb=msbm6
\font\fivemsb=msbm5


\font\tenfrm=eufm10
\font\eightfrm=eufm8
\font\sevenfrm=eufm7
\font\sixfrm=eufm6
\font\fivefrm=eufm5

\font\tenfrb=eufb10
\font\eightfrb=eufb8
\font\sevenfrb=eufb7
\font\sixfrb=eufb6
\font\fivefrb=eufb5
\font\twelvescr=eusm10 at 12pt
\font\tenscr=eusm10
\font\eightscr=eusm8
\font\sevenscr=eusm7
\font\sixscr=eusm6
\font\fivescr=eusm5
\fi
\def\eightpoint{\def\rm{\fam0\eightrm}%
\textfont0=\eightrm
  \scriptfont0=\sixrm
  \scriptscriptfont0=\fiverm 
\textfont1=\eighti
  \scriptfont1=\sixi
  \scriptscriptfont1=\fivei 
\textfont2=\eightsy
  \scriptfont2=\sixsy
  \scriptscriptfont2=\fivesy 
\textfont3=\tenex
  \scriptfont3=\tenex
  \scriptscriptfont3=\tenex 
\textfont\itfam=\eightit
  \scriptfont\itfam=\sixit
  \scriptscriptfont\itfam=\fiveit 
  \def\it{\fam\itfam\eightit}%
\textfont\slfam=\eightsl
  \scriptfont\slfam=\sixsl
  \scriptscriptfont\slfam=\fivesl 
  \def\sl{\fam\slfam\eightsl}%
\textfont\ttfam=\eighttt
  \def\tt{\fam\ttfam\eighttt}%
\textfont\bffam=\eightbf
  \scriptfont\bffam=\sixbf
  \scriptscriptfont\bffam=\fivebf
  \def\bf{\fam\bffam\eightbf}%
\textfont\frakfam=\eightfrm
  \scriptfont\frakfam=\sixfrm
  \scriptscriptfont\frakfam=\fivefrm
  \def\frak{\fam\frakfam\eightfrm}%
\textfont\frakbfam=\eightfrb
  \scriptfont\frakbfam=\sixfrb
  \scriptscriptfont\frakbfam=\fivefrb
  \def\bfrak{\fam\frakbfam\eightfrb}%
\textfont\scriptfam=\eightscr
  \scriptfont\scriptfam=\sixscr
  \scriptscriptfont\scriptfam=\fivescr
  \def\script{\fam\scriptfam\eightscr}%
\textfont\msbfam=\eightmsb
  \scriptfont\msbfam=\sixmsb
  \scriptscriptfont\msbfam=\fivemsb
  \def\bb{\fam\msbfam\eightmsb}%
\textfont\truecmr=\eighttruecmr
  \scriptfont\truecmr=\sixtruecmr
  \scriptscriptfont\truecmr=\fivetruecmr
  \def\truerm{\fam\truecmr\eighttruecmr}%
\textfont\truecmsy=\eighttruecmsy
  \scriptfont\truecmsy=\sixtruecmsy
  \scriptscriptfont\truecmsy=\fivetruecmsy
\tt \ttglue=.5em plus.25em minus.15em 
\normalbaselineskip=9pt
\setbox\strutbox=\hbox{\vrule height7pt depth2pt width0pt}%
\normalbaselines
\rm
}

\def\tenpoint{\def\rm{\fam0\tenrm}%
\textfont0=\tenrm
  \scriptfont0=\sevenrm
  \scriptscriptfont0=\fiverm 
\textfont1=\teni
  \scriptfont1=\seveni
  \scriptscriptfont1=\fivei 
\textfont2=\tensy
  \scriptfont2=\sevensy
  \scriptscriptfont2=\fivesy 
\textfont3=\tenex
  \scriptfont3=\tenex
  \scriptscriptfont3=\tenex 
\textfont\itfam=\tenit
  \scriptfont\itfam=\sevenit
  \scriptscriptfont\itfam=\fiveit 
  \def\it{\fam\itfam\tenit}%
\textfont\slfam=\tensl
  \scriptfont\slfam=\sevensl
  \scriptscriptfont\slfam=\fivesl 
  \def\sl{\fam\slfam\tensl}%
\textfont\ttfam=\tentt
  \def\tt{\fam\ttfam\tentt}%
\textfont\bffam=\tenbf
  \scriptfont\bffam=\sevenbf
  \scriptscriptfont\bffam=\fivebf
  \def\bf{\fam\bffam\tenbf}%
\textfont\frakfam=\tenfrm
  \scriptfont\frakfam=\sevenfrm
  \scriptscriptfont\frakfam=\fivefrm
  \def\frak{\fam\frakfam\tenfrm}%
\textfont\frakbfam=\tenfrb
  \scriptfont\frakbfam=\sevenfrb
  \scriptscriptfont\frakbfam=\fivefrb
  \def\bfrak{\fam\frakbfam\tenfrb}%
\textfont\scriptfam=\tenscr
  \scriptfont\scriptfam=\sevenscr
  \scriptscriptfont\scriptfam=\fivescr
  \def\script{\fam\scriptfam\tenscr}%
\textfont\msbfam=\tenmsb
  \scriptfont\msbfam=\sevenmsb
  \scriptscriptfont\msbfam=\fivemsb
  \def\bb{\fam\msbfam\tenmsb}%
\textfont\truecmr=\tentruecmr
  \scriptfont\truecmr=\seventruecmr
  \scriptscriptfont\truecmr=\fivetruecmr
  \def\truerm{\fam\truecmr\tentruecmr}%
\textfont\truecmsy=\tentruecmsy
  \scriptfont\truecmsy=\seventruecmsy
  \scriptscriptfont\truecmsy=\fivetruecmsy
\tt \ttglue=.5em plus.25em minus.15em 
\normalbaselineskip=12pt
\setbox\strutbox=\hbox{\vrule height8.5pt depth3.5pt width0pt}%
\normalbaselines
\rm
}

\def\twelvepoint{\def\rm{\fam0\twelverm}%
\textfont0=\twelverm
  \scriptfont0=\tenrm
  \scriptscriptfont0=\eightrm 
\textfont1=\twelvei
  \scriptfont1=\teni
  \scriptscriptfont1=\eighti 
\textfont2=\twelvesy
  \scriptfont2=\tensy
  \scriptscriptfont2=\eightsy 
\textfont3=\twelveex
  \scriptfont3=\twelveex
  \scriptscriptfont3=\twelveex 
\textfont\itfam=\twelveit
  \scriptfont\itfam=\tenit
  \scriptscriptfont\itfam=\eightit 
  \def\it{\fam\itfam\twelveit}%
\textfont\slfam=\twelvesl
  \scriptfont\slfam=\tensl
  \scriptscriptfont\slfam=\eightsl 
  \def\sl{\fam\slfam\twelvesl}%
\textfont\ttfam=\twelvett
  \def\tt{\fam\ttfam\twelvett}%
\textfont\bffam=\twelvebf
  \scriptfont\bffam=\tenbf
  \scriptscriptfont\bffam=\eightbf
  \def\bf{\fam\bffam\twelvebf}%
\textfont\scriptfam=\twelvescr
  \scriptfont\scriptfam=\tenscr
  \scriptscriptfont\scriptfam=\eightscr
  \def\script{\fam\scriptfam\twelvescr}%
\textfont\msbfam=\twelvemsb
  \scriptfont\msbfam=\tenmsb
  \scriptscriptfont\msbfam=\eightmsb
  \def\bb{\fam\msbfam\twelvemsb}%
\textfont\truecmr=\twelvetruecmr
  \scriptfont\truecmr=\tentruecmr
  \scriptscriptfont\truecmr=\eighttruecmr
  \def\truerm{\fam\truecmr\twelvetruecmr}%
\textfont\truecmsy=\twelvetruecmsy
  \scriptfont\truecmsy=\tentruecmsy
  \scriptscriptfont\truecmsy=\eighttruecmsy
\tt \ttglue=.5em plus.25em minus.15em 
\setbox\strutbox=\hbox{\vrule height7pt depth2pt width0pt}%
\normalbaselineskip=15pt
\normalbaselines
\rm
}
%
\fontdimen16\tensy=2.7pt
\fontdimen13\tensy=4.3pt
\fontdimen17\tensy=2.7pt
\fontdimen14\tensy=4.3pt
\fontdimen18\tensy=4.3pt
\fontdimen16\eightsy=2.7pt
\fontdimen13\eightsy=4.3pt
\fontdimen17\eightsy=2.7pt
\fontdimen14\eightsy=4.3pt
\fontdimen18\sevensy=4.3pt
\fontdimen16\sevensy=1.8pt
\fontdimen13\sevensy=4.3pt
\fontdimen17\sevensy=2.7pt
\fontdimen14\sevensy=4.3pt
\fontdimen18\sevensy=4.3pt
%
\def\hexnumber#1{\ifcase#1 0\or1\or2\or3\or4\or5\or6\or7\or8\or9\or
 A\or B\or C\or D\or E\or F\fi}
\mathcode`\=="3\hexnumber\truecmr3D
\mathchardef\not="3\hexnumber\truecmsy36
\mathcode`\+="2\hexnumber\truecmr2B
\mathcode`\(="4\hexnumber\truecmr28
\mathcode`\)="5\hexnumber\truecmr29
\mathcode`\!="5\hexnumber\truecmr21
\mathcode`\(="4\hexnumber\truecmr28
\mathcode`\)="5\hexnumber\truecmr29

\def\bar{\mathaccent"0\hexnumber\truecmr16 }

\def\Phi{\mathchar"0\hexnumber\truecmr08 }
\def\Gamma {\mathchar"0\hexnumber\truecmr00 }
\def\Delta {\mathchar"0\hexnumber\truecmr01 }
\def\Theta {\mathchar"0\hexnumber\truecmr02 }
\def\Lambda{\mathchar"0\hexnumber\truecmr03 }
\def\Xi {\mathchar"0\hexnumber\truecmr04 }
\def\Pi{\mathchar"0\hexnumber\truecmr05 }
\def\Sigma{\mathchar"0\hexnumber\truecmr06 }
\def\Upsilon {\mathchar"0\hexnumber\truecmr07 }
\def\Phi {\mathchar"0\hexnumber\truecmr08 }
\def\Psi {\mathchar"0\hexnumber\truecmr09 }
\def\Omega{\mathchar"0\hexnumber\truecmr0A }
\newcount\EQNcount \EQNcount=1
\newcount\CLAIMcount \CLAIMcount=1
\newcount\SECTIONcount \SECTIONcount=0
\newcount\SUBSECTIONcount \SUBSECTIONcount=1
\def\ifff#1#2#3{\ifundefined{#1#2}%
\expandafter\xdef\csname #1#2\endcsname{#3}\else%
\immediate\write16{!!!!!doubly defined #1,#2}\fi}
\def\NEWDEF#1#2#3{\ifff{#1}{#2}{#3}}
\def\actualnumber{\number\SECTIONcount}
\def\EQ#1{\lmargin{#1}\eqno\tageck{#1}}
\def\NR#1{&\lmargin{#1}\tageck{#1}\cr}  
\def\tageck#1{\lmargin{#1}({\rm \actualnumber}.\number\EQNcount)
 \NEWDEF{e}{#1}{(\actualnumber.\number\EQNcount)}
\global\advance\EQNcount by 1
}
\def\SECT#1#2{\lmargin{#1}\SECTION{#2}%
\NEWDEF{s}{#1}{\actualnumber}%
}

\def\CLAIM#1#2#3\par{
\vskip.1in\medbreak\noindent
{\lmargin{#2}\bf #1\ \actualnumber.\number\CLAIMcount.} {\sl #3}\par
\NEWDEF{c}{#2}{#1\ \actualnumber.\number\CLAIMcount}
\global\advance\CLAIMcount by 1
\ifdim\lastskip<\medskipamount
\removelastskip\penalty55\medskip\fi}
\def\CLAIMNONR #1#2#3\par{
\vskip.1in\medbreak\noindent
{\lmargin{#2}\bf #1.} {\sl #3}\par
\NEWDEF{c}{#2}{#1}
\global\advance\CLAIMcount by 1
\ifdim\lastskip<\medskipamount
\removelastskip\penalty55\medskip\fi}
\def\SECTION#1{\vskip0pt plus.2\vsize\penalty-75
    \vskip0pt plus -.2\vsize
    \global\advance\SECTIONcount by 1
    \beforesectionskip\noindent
{\sectionsize\sectiontype \actualnumber.\ #1}
    \EQNcount=1
    \CLAIMcount=1
    \SUBSECTIONcount=1
    \nobreak\sectionskip\noindent}
\def\SECTIONNONR#1{\vskip0pt plus.3\vsize\penalty-75
    \vskip0pt plus -.3\vsize
    \global\advance\SECTIONcount by 1
    \beforesectionskip\noindent
{\sectionsize\sectiontype  #1}
     \EQNcount=1
     \CLAIMcount=1
     \SUBSECTIONcount=1
     \nobreak\sectionskip\noindent}
\def\SUBSECTION#1{\vskip0pt plus.2\vsize\penalty-75%
    \vskip0pt plus -.2\vsize%
    \beforesectionskip\noindent%
{\subsectionsize\subsectiontype \actualnumber.\number\SUBSECTIONcount.\ #1}
    \global\advance\SUBSECTIONcount by 1
    \nobreak\sectionskip\noindent}
\def\SUBSECTIONNONR#1\par{\vskip0pt plus.2\vsize\penalty-75
    \vskip0pt plus -.2\vsize
\beforesectionskip\noindent
{\subsectionsize\subsectiontype #1}
    \nobreak\sectionskip\noindent\noindent}
\def\ifundefined#1{\expandafter\ifx\csname#1\endcsname\relax}
\def\equ#1{\ifundefined{e#1}$\spadesuit$#1\else\csname e#1\endcsname\fi}
\def\clm#1{\ifundefined{c#1}$\spadesuit$#1\else\csname c#1\endcsname\fi}
\def\sec#1{\ifundefined{s#1}$\spadesuit$#1
\else Section \csname s#1\endcsname\fi}
\def\lab#1#2{\ifundefined{#1#2}$\spadesuit$#2\else\csname #1#2\endcsname\fi}
\def\fig#1{\ifundefined{fig#1}$\spadesuit$#1\else\csname fig#1\endcsname\fi}
\let\endarg=\par
\def\finish{\def\endarg{\par\endgroup}}
\def\start{\endarg\begingroup}

 \def\beginFROM{\start\parskip=0pt\vskip\baselineskip
\def\finish{\def\endarg{\egroup\par\endgroup}}
  \vbox\bgroup\obeylines\eightpoint\em\finish}

\def\ABSTRACT#1\par{
\vskip 1in {\noindent\sectionsize\sectiontype Abstract.} #1 \par}

\def\TODAY{\number\day~\ifcase\month\or January \or February \or March \or
April \or May \or June
\or July \or August \or September \or October \or November \or December \fi
\number\year\timecount=\number\time
\divide\timecount by 60
}
\newcount\timecount
\def\DRAFT{\def\lmargin##1{\strut\vadjust{\kern-\strutdepth
\vtop to \strutdepth{
\baselineskip\strutdepth\vss\rlap{\kern-1.2 truecm\eightpoint{##1}}}}}
\font\footfont=cmti7
\footline={{\footfont \hfil File:\jobname, \TODAY,  \number\timecount h}}
}
\newbox\strutboxJPE
\setbox\strutboxJPE=\hbox{\strut}
\def\subitem#1#2\par{\vskip\baselineskip\vskip-\ht\strutboxJPE{\item{#1}#2}}
\gdef\strutdepth{\dp\strutbox}
\def\lmargin#1{}
\def\hexnumber#1{\ifcase#1 0\or1\or2\or3\or4\or5\or6\or7\or8\or9\or
 A\or B\or C\or D\or E\or F\fi}
\textfont\msbfam=\tenmsb
\scriptfont\msbfam=\sevenmsb
\scriptscriptfont\msbfam=\fivemsb
\mathchardef\varkappa="0\hexnumber\msbfam7B%
\newcount\FIGUREcount \FIGUREcount=0
\newdimen\figcenter
\def\definefigure#1{\global\advance\FIGUREcount by 1%
\NEWDEF{fig}{#1}{Fig.\ \number\FIGUREcount}
\immediate\write16{  FIG \number\FIGUREcount : #1}}
\def\figure#1#2#3#4\cr{\null%
\definefigure{#1}
{\goodbreak\figcenter=\hsize\relax
\advance\figcenter by -#3truecm
\divide\figcenter by 2
\midinsert\vskip #2truecm\noindent\hskip\figcenter
\includegraphics{#1}\vskip 0.8truecm\noindent \vbox{\eightpoint\noindent
{\bf\fig{#1}}: #4}\endinsert}}
\def\figurewithtex#1#2#3#4#5\cr{\null%
\definefigure{#1}
{\goodbreak\figcenter=\hsize\relax
\advance\figcenter by -#4truecm
\divide\figcenter by 2
\midinsert\vskip #3truecm\noindent\hskip\figcenter
\includegraphics{#1}{\hskip\texpscorrection\input #2 }\vskip 0.8truecm\noindent \vbox{\eightpoint\noindent
{\bf\fig{#1}}: #5}\endinsert}}
\def\figurewithtexplus#1#2#3#4#5#6\cr{\null%
\definefigure{#1}
{\goodbreak\figcenter=\hsize\relax
\advance\figcenter by -#4truecm
\divide\figcenter by 2
\midinsert\vskip #3truecm\noindent\hskip\figcenter
\includegraphics{#1}{\hskip\texpscorrection\input #2 }\vskip #5truecm\noindent \vbox{\eightpoint\noindent
{\bf\fig{#1}}: #6}\endinsert}}
\catcode`@=11
\def\footnote#1{\let\@sf\empty 
  \ifhmode\edef\@sf{\spacefactor\the\spacefactor}\/\fi
  #1\@sf\vfootnote{#1}}
\def\vfootnote#1{\insert\footins\bgroup\eightpoint
  \interlinepenalty\interfootnotelinepenalty
  \splittopskip\ht\strutbox 
  \splitmaxdepth\dp\strutbox \floatingpenalty\@MM
  \leftskip\z@skip \rightskip\z@skip \spaceskip\z@skip \xspaceskip\z@skip
  \textindent{#1}\footstrut\futurelet\next\fo@t}
\def\fo@t{\ifcat\bgroup\noexpand\next \let\next\f@@t
  \else\let\next\f@t\fi \next}
\def\f@@t{\bgroup\aftergroup\@foot\let\next}
\def\f@t#1{#1\@foot}
\def\@foot{\strut\egroup}
\def\footstrut{\vbox to\splittopskip{}}
\skip\footins=\bigskipamount 
\count\footins=1000 
\dimen\footins=8in 
\catcode`@=12 

\def\BB{{\script B}}
\def\CC{{\script C}}

\def\HALF{{\textstyle{1\over 2}}}

\def\QEDD{\hfill\smallskip
         \line{$\hfill{\vcenter{\vbox{\hrule height 0.2pt
	\hbox{\vrule width 0.2pt height 1.3ex \kern 1.3ex
		\vrule width 0.2pt}
	\hrule height 0.2pt}}}$
               \ \ \ \ \ \ }
         \bigskip}
\def\QED{$\hfill{\vcenter{\vbox{\hrule height 0.2pt
	\hbox{\vrule width 0.2pt height 1.3ex \kern 1.3ex
		\vrule width 0.2pt}
	\hrule height 0.2pt}}}$\bigskip}
\def\real{{\bf R}}

\def\complex{{\bf C}}

\def\Re{{\rm Re\,}}
\def\Im{{\rm Im\,}}
\def\PROOF{\medskip\noindent{\bf Proof.\ }}
\def\REMARK{\medskip\noindent{\bf Remark.\ }}
\def\LIKEREMARK#1{\medskip\noindent{\bf #1.\ }}
\normalbaselineskip=5.25mm
\baselineskip=5.25mm
\parskip=10pt
\beforesectionskipamount=24pt plus8pt minus8pt
\sectionskipamount=3pt plus1pt minus1pt
\def\em{\it}
\tenpoint
\null
\catcode`\@=11
\ifx\amstexloaded@\relax\catcode`\@=\active
\endinput\fi
\catcode`\@=\active
\def\period{\unskip.\spacefactor3000 { }}
%
%
\newbox\noboxJPE
\newbox\byboxJPE
\newbox\paperboxJPE
\newbox\yrboxJPE
\newbox\jourboxJPE
\newbox\pagesboxJPE
\newbox\volboxJPE
\newbox\preprintboxJPE
\newbox\toappearboxJPE
\newbox\bookboxJPE
\newbox\bybookboxJPE
\newbox\publisherboxJPE
\newbox\inprintboxJPE
\def\refclearJPE{
   \setbox\noboxJPE=\null             \gdef\isnoJPE{F}
   \setbox\byboxJPE=\null             \gdef\isbyJPE{F}
   \setbox\paperboxJPE=\null          \gdef\ispaperJPE{F}
   \setbox\yrboxJPE=\null             \gdef\isyrJPE{F}
   \setbox\jourboxJPE=\null           \gdef\isjourJPE{F}
   \setbox\pagesboxJPE=\null          \gdef\ispagesJPE{F}
   \setbox\volboxJPE=\null            \gdef\isvolJPE{F}
   \setbox\preprintboxJPE=\null       \gdef\ispreprintJPE{F}
   \setbox\toappearboxJPE=\null       \gdef\istoappearJPE{F}
   \setbox\inprintboxJPE=\null        \gdef\isinprintJPE{F}
   \setbox\bookboxJPE=\null           \gdef\isbookJPE{F}  \gdef\isinbookJPE{F}
     
   \setbox\bybookboxJPE=\null         \gdef\isbybookJPE{F}
   \setbox\publisherboxJPE=\null      \gdef\ispublisherJPE{F}
}
\def\widestlabel#1{\setbox0=\hbox{#1\enspace}\refindent=\wd0\relax}
\def\ref{\refclearJPE}
\def\no#1{\gdef\isnoJPE{T}\setbox\noboxJPE=\hbox{#1}}
\def\by#1{\gdef\isbyJPE{T}\setbox\byboxJPE=\hbox{#1}}
\def\paper#1{\gdef\ispaperJPE{T}\setbox\paperboxJPE=\hbox{#1}}
\def\yr#1{\gdef\isyrJPE{T}\setbox\yrboxJPE=\hbox{#1}}
\def\jour#1{\gdef\isjourJPE{T}\setbox\jourboxJPE=\hbox{#1}}
\def\pages#1{\gdef\ispagesJPE{T}\setbox\pagesboxJPE=\hbox{#1}}
\def\vol#1{\gdef\isvolJPE{T}\setbox\volboxJPE=\hbox{\bf #1}}
\def\preprint#1{\gdef
\ispreprintJPE{T}\setbox\preprintboxJPE=\hbox{#1}}

\def\book#1{\gdef\isbookJPE{T}\setbox\bookboxJPE=\hbox{\em #1}}
\def\publisher#1{\gdef
\ispublisherJPE{T}\setbox\publisherboxJPE=\hbox{#1}}
\def\inbook#1{\gdef\isinbookJPE{T}\setbox\bookboxJPE=\hbox{\em #1}}
\def\bybook#1{\gdef\isbybookJPE{T}\setbox\bybookboxJPE=\hbox{#1}}
\newdimen\refindent
\refindent=5em
\def\endref{\sfcode`.=1000
 \if T\isnoJPE
\hangindent\refindent\hangafter=1
      \noindent\hbox to\refindent{[\unhbox\noboxJPE\unskip]\hss}\ignorespaces
     \else  \noindent    \fi
 \if T\isbyJPE    \unhbox\byboxJPE\unskip: \fi
 \if T\ispaperJPE \unhbox\paperboxJPE\unskip\period \fi
 \if T\isbookJPE {\it\unhbox\bookboxJPE\unskip}\if T\ispublisherJPE, \else.
\fi\fi
 \if T\isinbookJPE In {\it\unhbox\bookboxJPE\unskip}\if T\isbybookJPE,
\else\period \fi\fi
 \if T\isbybookJPE  (\unhbox\bybookboxJPE\unskip)\period \fi
 \if T\ispublisherJPE \unhbox\publisherboxJPE\unskip \if T\isjourJPE, \else\if
T\isyrJPE \  \else\period \fi\fi\fi
 \if T\istoappearJPE (To appear)\period \fi
 \if T\ispreprintJPE Pre\-print\period \fi
 \if T\isjourJPE    \unhbox\jourboxJPE\unskip\ \fi
 \if T\isvolJPE     \unhbox\volboxJPE\unskip\if T\ispagesJPE, \else\ \fi\fi
 \if T\ispagesJPE   \unhbox\pagesboxJPE\unskip\  \fi
 \if T\isyrJPE      (\unhbox\yrboxJPE\unskip)\period \fi
 \if T\isinprintJPE (in print)\period \fi
\filbreak
}
\normalbaselineskip=12pt
\baselineskip=12pt
\parskip=0pt
\parindent=22.222pt
\beforesectionskipamount=24pt plus0pt minus6pt
\sectionskipamount=7pt plus3pt minus0pt
\overfullrule=0pt
\hfuzz=2pt
\nopagenumbers
\headline={\ifnum\pageno>1 {\hss\tenrm-\ \folio\ -\hss} \else
{\hfill}\fi}
\if F\PSfonts
\font\titlefont=cmbx10 scaled\magstep2

\font\toplinefont=cmr10
\font\pagenumberfont=cmr10
\let\tenpoint=\rm
\else
\font\titlefont=ptmb at 14 pt

\font\toplinefont=cmcsc10
\font\pagenumberfont=ptmb at 10pt
\fi
\newdimen\itemindent\itemindent=1.5em

\def\textindent#1{\indent\llap{#1\enspace}\ignorespaces}
\def\item{\par\noindent
\hangindent\itemindent\hangafter=1\relax
\setitemmark}
\def\setitemindent#1{\setbox0=\hbox{\ignorespaces#1\unskip\enspace}%
\itemindent=\wd0\relax
\message{|\string\setitemindent: Mark width modified to hold
         |`\string#1' plus an \string\enspace\space gap. }%
}
\def\setitemmark#1{\checkitemmark{#1}%
\hbox to\itemindent{\hss#1\enspace}\ignorespaces}
\def\checkitemmark#1{\setbox0=\hbox{\enspace#1}%
\ifdim\wd0>\itemindent
   \message{|\string\item: Your mark `\string#1' is too wide. }%
\fi}
\def\SECTION#1{\vskip0pt plus.2\vsize\penalty-75
    \vskip0pt plus -.2\vsize
    \global\advance\SECTIONcount by 1
    \beforesectionskip\noindent
{\sectionsize\sectiontype \actualnumber.\ #1}
    \EQNcount=1
    \CLAIMcount=1
    \SUBSECTIONcount=1
    \nobreak\sectionskip\noindent}
\catcode`\@=11
\def\@checkfirstline{\ifundefined{nonemptydefs}
\gdef\@haddefs{F}
\immediate\write16{--------Recreating defs.lst}
\else
\gdef\@haddefs{T}
\immediate\write16{--------Using old defs.lst}
\fi
}
\def\@definereally#1#2#3{\expandafter\xdef\csname #1#2\endcsname{#3}\relax}
\def\@writetofile#1#2#3{%
\write\@defs{\string\expandafter\string\xdef\string\csname}\relax
\write\@defs{\string #1#2\string\endcsname{\csname #1#2\endcsname}}}
\gdef\@haddefs{T}
\gdef\@neednewrun{T}
\gdef\@erasedefs{F}
\gdef\@doubly{F}
\newread\@defs
\openin\@defs defs.lst
\ifeof\@defs
\gdef\@haddefs{F}
\immediate\write16{*********Creating defs.lst}
\else
\closein\@defs
\input defs.lst
\@checkfirstline
\fi
\openout\@defs=defs.lst
\write\@defs{\string\def\string\nonemptydefs{}}
\def\@phaseone(#1,#2,#3){\ifundefined{#1#2}
\@definereally{#1}{#2}{#3}
\@writetofile{#1}{#2}{#3}
\else
 \write16{!!!!!doubly defined #1,#2}
 \gdef\@doubly{T}
\fi
}
\def\@phasetwo(#1,#2,#3){\@undefineda{#1#2}
\else
\edef\@firstarg{\csname #1#2\endcsname}
\edef\@secondarg{#3}
\ifnum \@stringcompare{\@firstarg}{\@secondarg} = 0
\else
\immediate\write16{ definition of #1 #2 changed : \@firstarg  --->\@secondarg}
\gdef\@neednewrun{T}
\gdef\@erasedefs{T}
\fi
\fi
\@definereally{#1}{#2}{#3}
\@writetofile{#1}{#2}{#3}}
\def\@undefineda#1{\expandafter\ifx\csname#1\endcsname\relax
{\gdef\@neednewrun{T}
 \immediate\write16{*********** used #1 which was never defined }}}
\if F\@haddefs
 \def\NEWDEF#1#2#3{\@phaseone({#1},{#2},{#3})}
\else
 \def\@neednewrun{F}
 \let\ifundefined=\@undefineda
 \def\NEWDEF#1#2#3{\@phasetwo({#1},{#2},{#3})}
\fi
\def\bye{\if T\@erasedefs
 \write16{**** I suspect doubly defined tokens ****}
 \write16{**** I erase defs.lst now ***********}
 \openout\@defs=defs.lst
\fi
\if T\@doubly
 \write16{**** There are doubly defined tokens *****}
 \write16{**** You have to correct the TeX file and rerun *****}
\fi
\if T\@neednewrun
 \write16{*********NEED ANOTHER RUN************}
\fi
\end}
\def\@stringcompare#1#2{\expandafter\@strcompa#1|===#2|===}
\def\@strcompa#1===#2==={\expandafter\@strcompb#2===#1===}
\def\@strcompb#1===#2==={\@compcont#1\\#2\\}
\def\@compcont#1#2\\#3#4\\{
\csname @comp
 \if #1#3\if #1|same\else contt\fi
   \else diff\fi
  \endcsname #2\\#4\\}
\def\@compcontt#1#2\\#3#4\\{\@strcompb#1#2===#3#4===}
\def\@compsame#1\\#2\\{0}
\def\@compdiff#1\\#2\\{1}
\catcode`\@=\active
\let\epsilon=\varepsilon
\def\GG{G}

\def\KK{{\cal K}}
\def\QQ{{\cal Q}}
\def\d#1{{\rm d}#1\,}
\def\kd#1{\kern-0.8em\d{#1}}
\def\tG{G}
\def\J{J_\xi}

\def\CC{{\cal C}}
\def\Cgreen{C(\beta )}
\let\truett=\tt
\fontdimen3\tentt=2pt\fontdimen4\tentt=2pt
\def\tt{\hfill\break\null\kern -2truecm\truett **** }
\def\uncheckedifundefined#1{\expandafter\ifx\csname#1\endcsname\relax}
\def\UNCHECKEDNEWDEF#1#2#3{\expandafter\xdef\csname #1#2\endcsname{#3}}%
\newcount\Ccount \Ccount=0
\def\C#1{\lmargin{C#1}
{\uncheckedifundefined{w#1}%
\global\advance\Ccount by 1}%
\UNCHECKEDNEWDEF{w}{#1}{C_{\number\Ccount}%
\fi}
\csname w#1\endcsname%
}
\setitemindent{iii)}
{\titlefont{\centerline{Proof of the Marginal Stability}}}
\vskip 0.5truecm
{\titlefont{\centerline{Bound for the Swift-Hohenberg Equation}}}
\vskip 0.5truecm
{\titlefont{\centerline{and Related Equations}}}
\vskip 0.5truecm
{\it{\centerline{ P. Collet${}^{1}$ and J.-P. Eckmann${}^{2,3}$}}}
\vskip 0.3truecm
{\eightpoint
\centerline{${}^1$Centre de Physique Th\'eorique, Laboratoire CNRS UMR
7644,
Ecole Polytechnique, F-91128 Palaiseau Cedex, France}
\centerline{${}^2$D\'ept.~de Physique Th\'eorique, Universit\'e de Gen\`eve,
CH-1211 Gen\`eve 4, Switzerland}
\centerline{${}^3$Section de Math\'ematiques, Universit\'e de Gen\`eve,
CH-1211 Gen\`eve 4, Switzerland}
}
\vskip 0.5truecm\headline
{\ifnum\pageno>1 {\toplinefont Marginal Stability Bound}
\hfill{\pagenumberfont\folio}\fi}
\narrower
{\eightpoint\baselineskip 9pt
\LIKEREMARK{Abstract}We prove that if the initial condition of the
Swift-Hohenberg equation 
$$\partial _t u(x,t)=\bigl(\epsilon
^2-(1+\partial_ x^2)^2\bigr) u(x,t)
-u^3(x,t)
$$ 
is bounded in modulus by $Ce^{-\beta x }$ as $x\to+\infty $, the
solution cannot propagate to the right with a speed greater than
$$
\sup_{0<\gamma\le\beta }\gamma^{-1}(\epsilon ^2+4\gamma^2+8\gamma^4)~.
$$
This settles a long-standing conjecture about the possible asymptotic
propagation speed of the Swift-Hohenberg equation. The proof does not
use the maximum principle and is simple enough to  generalize
easily to other equations. We illustrate this with an example of a
modified Ginzburg-Landau equation, where the minimal speed is not
determined by the linearization alone.
}\par
\advance\leftskip-\parindent
\advance\rightskip-\parindent
\SECTION{Introduction}The marginal stability conjecture deals with the
possible propagation 
speed of solutions of dissipative partial differential equations. It
was formulated in the late 1970's by several authors.
Its clearest form is obtained for the Ginzburg-Landau equation
$$
\partial _t u(x,t)\,=\,\partial _x^2 u(x,t)+ u(x,t)-u^3(x,t)~,
\EQ{GL}
$$
where $u:\real\times\real^+\to\real$. When the initial data have
compact support, then the solution cannot propagate with a speed
faster than some critical speed $c$, which happens to be 2 for this
example.
The number 2 can be understood as follows. One writes
$u(x,t)=v(x-ct)$, and looks for a solution
of \equ{GL} expressed for $v$:
$$
0\,=\,\partial _x^2v +c\partial_x v +v -v^3~.
\EQ{GLc} 
$$
If one makes the assumption that $v(\xi)= \C{v}e^{-\beta \xi}$ as
$\xi\to+\infty $, one finds obviously that $\beta $ and $c$ should be
related through the equation 
$$
0\,=\,\beta ^2-\beta c+1~,
\EQ{betac} 
$$
since the non-linear term is irrelevant at $\xi=\infty $ in this case.
For fixed $\beta $ we clearly find $c=(\beta ^2+1)/\beta $, and since
functions which are (in absolute value) bounded by $C\exp(-\beta x) $ 
are also bounded by
$C'\exp(-\beta ' x)$ for $0<\beta '<\beta $ one finds in this case an
upper bound 
$$
c_\beta^{\rm GL} \,=\,\inf_{0<\gamma<\beta}{\gamma ^2+1\over\gamma } ~,
\EQ{glbound} 
$$
and this is equal to 2 for $\beta \ge 1$.
Using the maximum principle for parabolic PDE's, Aronson and
Weinberger were able to show [AW] that no positive solution starting
from initial conditions with compact support can move faster than the
speed $c_2^{\rm GL}=2$. Using essentially the same argument, it was also shown
in [CE] that if the initial condition decays like $e^{-\beta x}$ with
$\beta < 1$, then the solution cannot move faster than $c_\beta^{\rm GL}$.
However, in cases where the maximum principle does not apply, such as
in \equ{SH}, the maximum possible speed was only conjectured, and
tested numerically,
but no rigorous result was obtained, see {\it e.g.}, [LMK, DL,
BBDKL].

In a somewhat different direction,
there is the important, and difficult, issue on whether there
is actually a solution moving with the maximal allowed velocity.
In general, its realization depends
on the details of the nonlinearity, and this question has been
extensively discussed in the literature [AW, B, DL, BBDKL, vS]. It
will not be treated here.

The main result of our paper is an {\em upper bound} on the speed of
propagation of solutions to the Swift-Hohenberg equation
$$
\partial _t u \,=\,\bigl(\epsilon -(1+\partial_x^2)^2\bigr )u -u^3~.
\EQ{SH}
$$
The polynomial equation analogous to \equ{betac} turns out to be
$$
0\,=\,\epsilon ^2 +4\beta ^2+8\beta ^4 -c\beta ~,
\EQ{newcbeta} 
$$
and we define in this case
$$
c_\beta\,=\, c_\beta^{\rm SH}\,=\,\inf_{0<\gamma\le\beta } {\epsilon
^2 +4\gamma^2+8\gamma^4\over \gamma }~. 
\EQ{ccrit}
$$ 
The polynomial is an absolute maximum of the real part of
$P$, as we explain at the end of the introduction and in the Appendix.
This will be the minimal speed.\footnote{${}^*$}{While it looks
different from the standard discussion in [BBDKL], 
we explain in the Appendix that the two definitions coincide.
The current formulation has the advantage of being expressed in terms
of real variables, although the traveling wave in this case is
actually modulated [CE].} 
Our result can be expressed informally as follows: {\em If the initial
data for the problem are bounded in absolute value by 
$Ce^{-\beta x}$ as $x\to+\infty$ then the
solution cannot advance faster to the right than $c_\beta$ in the sense that
$$
\lim_{t\to \infty } u(x+ct,t)\,=\,0~,
$$
for all $c>c_\beta$. In particular, if the initial condition has
compact support, the above hypotheses are satisfied for any $\beta >0$
and we find an upper bound on the speed which is $c_*=
\inf_\beta c_\beta $: This is the absolute minimum of
$(\epsilon ^2 +4\beta ^2+8\beta ^4)/\beta $.
}

\REMARK The precise formulation is given in \clm{main}.

Before explaining the main steps of the proof
we note a well-known result, namely that if the initial condition
$u_0$ is
bounded in $\CC^3$, {\it i.e.},
$$
\max _{j=0,\dots,3}\,\sup_{x\in\real} |\partial _x^j u_0(x)|\,\le\,K~,
\EQ{bound0} 
$$
then there is a constant $L=L(K)$ such that for all $t>0$ one has
$$
\max _{j=0,\dots,3}\,\sup_{x\in\real} |\partial _x^j
u(x,t)|\,\le\,L(K)~.
\EQ{bound1} 
$$

The proof of the main result is really quite easy and consists of 3
steps:  
\item{i)}An a priori bound on the Green's function of the semigroup
generated by the linear part $\epsilon ^2-(1+\partial_x^2)^2$
of the Swift-Hohenberg equation.
\item{ii)}The observation that if the initial condition satisfies
$\lim_{x\to\infty }e^{\beta x}\partial_x^j u_0(x)=0$, for
$j=0,\dots,3$, then the same holds for $u(x,t)$. This is needed later
on to ensure that integration by parts does not produce boundary terms
at infinity.
\item{iii)}An energy-like estimate which shows that
$$
\lim_{t\to\infty}\int_{ct}^\infty \d{x} |u(x,t)|^2 e^{2\beta
(x-ct)}\,=\,0~,
$$
when $c>c_\beta$ (if it is finite at $t=0$, see below for details).
Thus, the solution is outrun by a frame moving with
speed $c>c_\beta $. While this is similar to what was observed in the
proofs where the maximum principle could be used, it has here a
quite different origin of dynamical nature.
\vskip 0.5truecm
In \sec{nonlin}, we consider the case of the Ginzburg-Landau equation
when the nonlinearity $u-u^3$ is replaced by a general function $f(u)$
with the properties $f(0)=0$, $0<f'(0)<\infty$ and
$\limsup_{z\to\infty } f(z)/z  <0$. In such a case, the bound
\equ{glbound} is replaced by
$$
c_\beta^{\rm GL'} \,=\,\inf_{0<\gamma<\beta}{\gamma ^2+\sup_u{f(u)\over
u}\over\gamma  }~.
$$
In the case of the Swift-Hohenberg equation the bound generalizes as
follows:
Assume the equation is
$$\partial _t u\,=\,\bigl(\epsilon ^2-(1+\partial _x^2)^2)\bigr)u+ f(u)~.
$$
Then we get for the maximal possible speed:
$$
c_\beta^{\rm SH'}\,=\,\inf_{0<\gamma\le\beta } {\epsilon ^2
+4\gamma^2+8\gamma^4+\sup_u{f(u)\over u}\over \gamma }~. 
$$

In an appendix, we show that the expression \equ{newcbeta} is
nothing but
$$
\sup_{k_\beta ^*}\Re P(z)|_{z=-\beta +ik^*_\beta }~,
$$
where the sup is over the solutions $k^*_\beta$ of
$$
{{\rm d}\Re P(-\beta +ik)\over \d{k}}\,=\,0~.
$$
We also show that these conditions are the same as those found in
[BBDKL].

Finally, it should be noted that the method is not restricted to
1-dimensional problems, and can also be applied to questions of grows
of ``bubbles'' in the 2-dimensional Swift-Hohenberg equation.
\SECT{pointwise}{A pointwise bound on the Green's function}Here we
bound the 
Green's function of the operator 
$\epsilon ^2 -(1+\partial _x^2)^2$ by a method which generalizes
immediately to other problems of similar type.
Let $P$ be a polynomial in $k$ which is of the form
$$
P(ik)\,=\,-a_{n} k^{n}+\sum_{m=0}^{n-1} a_{m}k^m~\,\equiv\, -a_{n} k^{n}+R(k)~,
$$
and assume $n$ even and $a_n>0$. 
(For the Swift-Hohenberg equation, $P(z)=\epsilon ^2-(1+z^2)^2$.)
Then the Green's function
$$
G_t(x)\,=\,\int\d{k} e^{ikx}e^{P(ik)t}~,
$$
satisfies:
\CLAIM{Lemma}{Green}Given $0<\beta <\infty $, there is a constant
$\Cgreen $ such that for 
all $t\in(0,1]$ one has the bound
$$
t^{1/n}|G_t(x)|e^{(\beta'+2t^{-1/n}) |x|} \,\le\,\Cgreen ~,
\EQ{green} 
$$
for all $\beta '\in[0,\beta ]$.

\REMARK This clearly also implies, for all $t\in(0,1]$ and all $\beta
'\in[0,\beta]$:
$$
\int \d{x} |G_t(x)|e^{\beta '|x|}\,\le\,\Cgreen ~,
\EQ{greenint}
$$ 
since $\int\d{x} |G_t(x)|e^{\beta '|x|}\le \Cgreen \int\d{x}t^{-1/n}
e^{-2|x|t^{-1/n}}\le \Cgreen $. 
\PROOF We will show the bound in the form
$$
t^{1/n}|G_t(zt^{1/n})|e^{\gamma t^{1/n}z}\,\le\,   \Cgreen ~,
\EQ{g1} 
$$
with $\gamma=\beta+2t^{-1/n}$,
and it clearly suffices to consider $z>0$.
Proving \equ{g1} is a straightforward calculation which is probably well-known.
Indeed, the l.h.s.~of \equ{g1} equals (without the absolute values)
$$
\eqalign{
&~~\int \d{k} t^{1/n} \exp\bigl(\gamma t^{1/n}z + ik t^{1/n}z -a_nk^{n}t +
R(k)t\bigr) \cr
\,&=\,
\int \d{\ell} \exp\bigl(\gamma t^{1/n}z+i\ell z -a_n \ell^n
+R(\ell t^{-1/n})t
\bigr)~.
}
$$
Since the integrand is an entire function in $\ell$ we can shift the
contour from $\ell$ to $\ell'=\ell-i\gamma t^{1/n}$ and the last expression
is seen to be equal to
$$
\int \d{\ell'} \exp\bigl(i\ell' z -a_n (\ell'+i\gamma t^{1/n})^n
+R(\ell' t^{-1/n}+i\gamma )t
\bigr)~.
$$
Note now that
$$
\eqalignno{
\bigl|&\exp\bigl(i\ell' z -a_n (\ell'+i\gamma t^{1/n})^n
+R(\ell' t^{-1/n}+i\gamma )t
\bigr )\bigr|\cr\,&=\,\bigl|\exp\bigl( -a_n (\ell'+i\gamma t^{1/n})^n
+R(\ell' t^{-1/n}+i\gamma )t
\bigr )\bigr|~,\NR{xyz}
}
$$
and for bounded $\beta $ and $t\in(0,1]$ we find that
$\gamma t^{1/n}=(\beta +2t^{-1/n})t^{1/n}\le \beta +2$, and hence \equ{xyz} is uniformly
integrable in $\ell'$, since $a_n>0$. 
The proof of \clm{Green} is complete.\QED

\SECTION{Exponential decay of solutions}In this section, we prove a
bound in the laboratory frame, showing that if the initial condition
goes exponentially to 0 then the solution at time $t$ goes to zero as
well, {\em with the same rate}.

\CLAIM{Theorem}{xbound}Assume that
$u_0 $ is bounded in $\CC^3$ and that
$$
\lim_{x\to\infty } e^{\beta x} \partial _x^j u_0(x) \,=\,0~,
\EQ{initial}
$$
for $j=0,\dots,3$ and some $\beta >0$. Then the solution $u(x,t)$ of
\equ{SH} with initial 
data $u_0$ satisfies
for all $t>0$:
$$
\lim_{x\to\infty } e^{\beta x} \partial _x^j u(x,t) \,=\,0~,
\EQ{final}
$$
for $j=0,\dots,3$.

\PROOF The proof is in steps of some (fixed) time $\tau _*$. 
We define first
$$
g_\xi(x)\,=\,1+e^{\beta (x-\xi)}~.
$$
The assumption means that $u_0$ satisfies \equ{bound0} for some $K$.
From \equ{initial}, and because $L(K)\ge K$, we conclude 
that there is a $\xi>0$ for which  
$$
\sup_{x\in\real } g_\xi(x) |\partial _x^j u_0(x)| \,\le\,2L(K)~,
\EQ{initial2}
$$
for $j=0,\dots,3$.
Note that we do not have any control on the size of $\xi$, but such a
control is not needed.

From \equ{bound0} we also conclude (see \equ{bound1}) that 
$$
\sup_{t\ge0}\sup _{x\in\real} | \partial _x^j u(x,t)| \,\le\, L(K)~,
\EQ{universalbound} 
$$
for $j=0,\dots,3$.

The crucial step in the proof of \clm{xbound} is 
\CLAIM{Lemma}{taustar}There are a $\tau _*>0$ and a $\rho$,
independent of $\xi$,
such that for $t\in[0,\tau _*]$ one has
$$
\sup_{j=0,\dots,3}\,\sup_{x\in\real } g_\xi(x) |\partial _x^j u(x,t)|
\,\le\,\rho~. 
\EQ{final2}
$$

\PROOF We use the estimates on the convolution kernel
$G_t$ associated with the semigroup $t\mapsto \exp\bigl(t(\epsilon
^2-(1+\partial _x^2)^2\bigr)$ which were proven in \sec{pointwise}.
One has 
$$
u_t\,=\,G_t\star u_0 - \int_0^t \d{s} G_{t-s}\star u_s^3~,
$$
where $u_s(x)=u(x,s)$.
We define $\BB_\xi$ as the space of uniformly continuous functions $f$
for which
$$
\|f\|_\xi\,=\,\sup_{x\in\real}  g_\xi(x)|f(x)| \,<\,\infty~.
$$ 
Using this quantity as a norm makes $\BB_\xi$ a Banach space.
Consider next the space $\KK=\KK_{\xi,\tau _*}=\CC^0([0,\tau
_*],\BB_\xi)$ of functions $h:(x,t)\mapsto h(x,t)$, with the norm 
$$
\|h\|_{\xi,\tau _*}\,=\,\sup_{t\in[0,\tau _*]}\|h(\cdot,t)\|_\xi~.
$$
This is again a Banach space.
For $v\in\KK$ we define the map $v\mapsto \QQ v$ by
$$
\bigl(\QQ v\bigr)(x,t)\,=\,\bigl(G_t\star u_0\bigr )(x) - \int_0^t \d{s} \bigl(G_{t-s}\star
v_s^3\bigr )(x)~.
\EQ{qmap} 
$$
Note that if $\QQ v = v$, then $v$ is a solution to \equ{SH} with
initial condition $u_0$. To find $v$, 
we will show that for sufficiently small $\tau _*>0$ the operator
$\QQ$ contracts a small ball of $\KK_{\xi,\tau _*} $ to itself.
The center of this ball is the function $(x,t)\mapsto 0$.

First we bound $\GG_t\star u_0$.
Note that from the definition of $g_\xi$ we find
$$
{g_\xi (x)\over g_\xi(y)}\,\le\,e^{\beta |x-y|}~,
$$
since for $x<y$ the quotient is bounded by 1 and for $x>y$ we have the
(very rough) bound $e^{\beta(x-y)} $. From \clm{Green}, we have
for all $t\in(0,1]$ and all $x\in\real$:
$$
|G_t(x)|e^{2\beta |x|} \,\le\,\Cgreen t^{-1/4}e^{-2|x|t^{-1/4}}~,
\EQ{green0} 
$$
and, clearly, $\Cgreen $ can be chosen the same value for all
smaller $\beta $.
Using this,
we find 
$$
\eqalign{
|\bigl(\GG_t\star u_0\bigr)(x)\,g_\xi (x)| \,&\le\,
\int \d{y}
|\tG_t(x-y) u_0(y)|\, g_\xi(y) {g_\xi(x)\over g_\xi(y)}\cr
 \,&\le\,\int \d{y}
|\tG_t(x-y) u_0(y)|\, g_\xi(y) e^{\beta |x-y|}\cr
 \,&\le\,\int \d{z}
|G_t(z) e^{\beta |z|}|\sup_{z'\in\real}|u_0(z') |\,g_\xi(z') \cr
\,&\le\,\Cgreen \sup_{z'\in\real}|u_0(z') |\,g_\xi(z')
~.\cr
} 
\EQ{e1}
$$
Combining these bounds with \equ{initial2}
we 
get
$$
|\bigl(\GG_t \star u_0\bigr )(x)\,g_\xi(x)|\,\le\, \C{3} L(K)~.
$$
In fact, we can do a little better in \equ{e1} by extracting a factor
of $e^{-\beta |x-y|}$. The last two lines in \equ{e1} are replaced by
$$
\eqalign{
|\bigl(&\GG_t\star u_0\bigr)(x)\,g_\xi (x)|\cr
 \,&\le\,\int \d{z}
|G_t(z) e^{2\beta |z|}|\sup_{y\in\real}|u_0(y) |\,g_\xi(y)e^{-\beta |x-y|} \cr
\,&\le\,C(2\beta) \sup_{y\in\real}|u_0(y) |\,g_\xi(y)e^{-\beta |x-y|}
~.\cr
}
\EQ{e1a}
$$
Since $|u_0(y) |\,g_\xi(y)$ is bounded and converges to 0 as
$y\to+\infty $, we conclude that the quantity in \equ{e1a} tends to 0
as $x\to+\infty $. 
Thus, we also have
$$
\lim_{x\to\infty}|\bigl(\GG_t\star u_0\bigr)(x)\,g_\xi (x)| \,=\,0~.
\EQ{homogeneous} 
$$

We next bound the non-linear term.
Assume $v\in\KK_{\xi,\tau _*}$ and $\|v\|_{\xi,\tau _*}<\rho$.
Then any power ($\ge1$) of $v$ is also in $\KK_{\xi,\tau _*}$ and one has a
bound of the form
$$
\|v^3\|_{\xi,\tau _*}\,\le\, \C{4} \rho^3~.
$$
Therefore, the method leading to \equ{e1} now yields
$$
\left |
\int_0^t\d{s}\bigl(\GG_{t-s}\star v_s^3\bigr )(x) g_\xi(x)
\right |
\,\le\,\C{5} \rho^3 t~,
$$
and if also $\|w\|_{\xi,\tau _*}<\rho$, then a variant of that method gives:
$$
\eqalign{
\bigg |
\int_0^t\d{s}\bigl( \GG_{t-s}&\star v_s^3\bigr )(x) g_\xi(x)
-
\int_0^t\d{s}\bigl(\GG_{t-s}\star w_s^3\bigr )(x) g_\xi(x)
\bigg | \cr
\,&\le\,\C{6} \rho^2 t
\sup _{s\in[0,t]} \sup_{x\in\real}
\bigl|v_s(x)-w_s(x)\bigr|\,g_\xi(x)~.\cr
}
$$
Taking the center of the ball at $(x,t)\mapsto 0$ and the radius
$\rho=2\C{3}K$ and then $\tau _*< \min\{(4\C{5}\rho^3)^{-1},(4\C{6}
\rho^2)^{-1}\}$, 
we have a contraction and hence a unique fixed point $v$ for $\QQ$. For
$j=1,2,3$, we use the same methods since we can push all derivatives
from the operator $\GG_t$ to the function $v$, because
$\GG_t\star$ is a convolution. The details are left to the reader.
The existence of this fixed point clearly shows \clm{taustar}.\QED

We come back to the proof of \clm{xbound}. We define
$$
\Gamma(t)\,=\,\limsup_{x\to\infty }  | u(x,t)g_\xi(x)|~.
$$
By assumption, we have $\Gamma(0)=0$ and by \clm{taustar} we have
$$
|u(x,t)|\,\le\,\rho/g_\xi(x)~,
$$
so that $\Gamma(t)\le\rho$ for $t\le\tau _*$. We now show it is actually 0
for those $t$.
Consider $\QQ$ as in \equ{qmap}. Note that 
$$
\eqalign{
\Gamma(t)\,&=\,\limsup_{x\to\infty}|u(x,t)g_\xi(x)| \cr
\,&=\,\limsup_{x\to\infty}g_\xi(x)|\bigl(G_t\star u_0\bigr )(x)|
+\limsup_{x\to\infty}
g_\xi(x)| \int_0^t \d{s} \bigl(G_{t-s}\star
u_s^3\bigr )(x)|~.
}
$$
The first term vanishes by \equ{e1a}.
Thus, $\Gamma$ only depends on the nonlinear
part. 
Using \equ{green0}, that part can be bounded as
$$
\eqalign{
g_\xi(x)\left |
\int_0^t\d{s} \GG_{t-s}\star u_s^3(x)\right |\,&\le\,
\int_0^t \d{s} \int\d{y}
{g_\xi(x)\over g_\xi^3(y)}
\,\left|\tG_{t-s}(x-y)\right|
\left | g_\xi(y) u_s(y)\right |^3 \cr
\,&\le\,
\int_0^t\d{s}\int \d{y}
\left|\tG_{t-s}(x-y)\right|e^{\beta |x-y|}\,\left | g_\xi(y)
u_s(y)\right |^3 \cr 
\,&\le\,
\Cgreen\int_0^t\d{s}\int \d{z}
(t-s)^{-1/4} e^{-2|z|(t-s)^{-1/4}}\cr
&\qquad\qquad\cdot
\left | g_\xi(x-z)
u_s(x-z)\right |^3~. \cr 
}
\EQ{nonlin} 
$$
We need an upper bound for the $\limsup_{x\to\infty }$ of this
expression.
Fix an $\epsilon >0$.
For $s\in[0,t]$, we can find an $\eta(s,\epsilon )>0$
such that
$$
\sup_{y\ge \eta(s,\epsilon )}|g_{\xi}(y)u_{s}(y)|\le \Gamma(s)+\epsilon\;.
$$ 
There is also a number $\zeta(\epsilon)>0$ such that for any $s\in[0,t]$:
$$
\int_{|z|>\zeta(\epsilon )}  \d{z} (t-s)^{-1/4}
e^{-2|z|(t-s)^{-1/4}}\le \epsilon~. 
$$
If $x>\zeta(\epsilon )+\eta(s,\epsilon )$, we have 
$$
\int \d{z} (t-s)^{-1/4} e^{-2|z|(t-s)^{-1/4}}
|g_{\xi}(x-z)u_{s}(x-z)|^{3}\le (\Gamma(s)+\epsilon)^{3}+ \rho^{3}\epsilon~,
$$
by \clm{taustar}. We cannot conclude directly by integration over $s$
because $\eta$ depends on $s$. However, $\eta(s,\epsilon )$ 
is finite for almost every
$s$ (in reality for every $s$). Therefore, we can find a finite  number
$\Theta(\epsilon)$ such that the set
$$
E(\epsilon)\,=\,\big\{ s\in[0,t]\;|\; \eta(s,\epsilon )\,>\,\Theta(\epsilon)\big\}
$$
has Lebesgue measure at most $\epsilon^{2}$ (note that $E(\epsilon )$ is
measurable). Therefore, if $x>\Theta(\epsilon)+\zeta(\epsilon)$ we have
$$
\eqalign{
\int_{0}^{t}&\d{s}\int \d{z} (t-s)^{-1/4} e^{-2|z|(t-s)^{-1/4}}
|g_{\xi}(x-z)u_{s}(x-z)|^{3}\cr
\,&=\,\int_{([0,t]\setminus E(\epsilon))\cup E(\epsilon )}\d{s}\int \d{z}
 (t-s)^{-1/4} e^{-2|z|(t-s)^{-1/4}}
|g_{\xi}(x-z)u_{s}(x-z)|^{3}\cr
\,&\le\,
\C{new} \int_{0}^{t}\d{s}\bigl((\Gamma(s)+\epsilon)^{3}+
\rho^3\epsilon\bigr )
+\C{new2} \rho^{3}\int_{E(\epsilon )}\d{s} 
(t-s)^{-1/4}~.
}
$$
The last integral is of order $\epsilon^{1/2}$ by the Schwarz inequality.
Since $\epsilon >0$ is arbitrary, we get
$$
\Gamma(t)\,\le\, \C{gamma} \int_0^t   
\d{s} \Gamma(s)^3~.
$$
Since $\Gamma$ is bounded by what we said above and $\Gamma(0)=0$, it
follows from 
Gronwall's lemma that $\Gamma(t)=0$ for $t\le\tau _*$. One then
repeats the argument for all consecutive intervals of length $\tau
_*$.
The proof of the
corresponding bounds on the derivatives is similar and is left to
the reader.\QED

\SECT{speed}{Bound on the speed}We define $\J$ by
$$
\J(t)\,=\,\int_\xi^\infty \kd{x} |u(x,t)|^2 e^{2\beta (x-\xi )}~,
\EQ{jdef} 
$$
where $u(x,t)$ is the solution of the Swift-Hohenberg equation.
The main result of this paper is 
\CLAIM{Theorem}{main}Let $u(x,t)$ be a solution of the Swift-Hohenberg
equation \equ{SH} for an initial condition $u_0(x)=u(x,0)$ which is in
$\BB$, which satisfies $J_0(0)<\infty $ for some $\beta >0$ and which
satisfies the
assumptions of \clm{xbound}. Then one has
$$
\lim_{t\to\infty } \int_{ct}^\infty \d{x}|u(x,t)|^2 e^{2\beta
(x-ct)}\,=\,0~,
\EQ{result} 
$$
for all $c > \bigl(\epsilon ^2+4\beta ^2+8\beta ^4\bigr)/\beta $.

\REMARK If one is willing to pay a price of slightly more complicated
formulations and proofs, one can omit the condition on $J_0(0)$ in
\clm{main}. One would then assume the pointwise bounds of \clm{xbound}
fore some $\beta >0$ and work throughout the proof with a $J_\xi(t)$
defined with some $\beta '<\beta $, but arbitrarily close to it,
since the condition on $c$ is open.

\PROOF We define $v_\xi (x,t)=u(x,t)e^{\beta (x-\xi )}$, so that
$\J(t)=\int_\xi ^\infty \d{x}|v_\xi (x,t)|^2$, and $v_\xi $ solves the
equation
$$
\partial_t v_\xi(x,t) \,=\,\epsilon ^2 v_\xi(x,t)  - \bigl(1+(\partial
_x-\beta )^2\bigr)^2 v_\xi(x,t)  -v_\xi ^3(x,t) e^{-2\beta (x-\xi )}~.
\EQ{SH2} 
$$
Since $u$ is real, the absolute values in the definition of $J_\xi(t)$
can be omitted.
Differentiating \equ{jdef} with respect to time, we get
$$
\eqalign{
\HALF \partial _t \J(t)\,&=\,
\int_\xi^\infty \kd{x} v_\xi(x,t)\partial _tv_\xi(x,t)~.
}
$$ 
Since $\xi$ is fixed throughout the calculation, we omit the index of
$v_\xi$. We also omit the arguments $(x,t)$. Note that by
\clm{xbound}, $\lim_{x\to\infty }\partial _x^jv_\xi(x,t)=0$, for
$j=0,\dots,3$, so that we can freely integrate by parts in the
following calculation. We find, using $\partial _x v=v'$:
$$
\eqalign{
\HALF \partial _t \J(t)\,&=\,
\int_\xi^\infty \kd{x}
v\biggl( \epsilon ^2 v   - \bigl(1+(\partial
_x-\beta )^2\bigr)^2 v  -v^3 e^{-2\beta (x-\xi )}\biggr)\cr
\,&=\,
\int_\xi^\infty \kd{x}
v\biggl( \epsilon ^2 v   - \bigl(1+\partial
_x^2-2\beta\partial _x +\beta ^2 \bigr)^2 v  -v^3 e^{-2\beta (x-\xi
)}\biggr)\cr
\,&=\,
\int_\xi^\infty \kd{x}
v\biggl( \epsilon ^2 v   - \partial _x^4v+4\beta \partial
_x^3v-2(1+3\beta ^2)\partial _x^2v+4\beta (1+\beta ^2)\partial_ xv
-(1+\beta ^2)^2v\cr&~~~~~~~~~~~~~~~~~~~~~ -v^3 e^{-2\beta (x-\xi
)}\biggr)\cr
\,&=\,
\int_\xi^\infty \kd{x}\biggl(
\bigl(\epsilon ^2-(1+\beta ^2)^2\bigr)v^2  - e^{-2\beta (x-\xi
)}v^4-2(1+3\beta ^2)vv''\cr
&~~~~~~~~~~~~~~~~~~~~~ +~v'v'''-4\beta v'v''\biggr)
\cr&~~~~~~~~~~~~~~~~~~~~~ +\bigl(vv'''-4\beta
vv''-2\beta (1+\beta ^2)v^2\bigr )\bigr|_{x=\xi,t}~.\cr  
}
$$
We integrate by parts some more and get
$$
\eqalign{
\HALF \partial _t \J(t)\,&=\,
\int_\xi^\infty \kd{x} 
\biggl(\bigl(\epsilon ^2-(1+\beta ^2)^2\bigr)v^2 - e^{-2\beta (x-\xi
)}v^4 -2(1+3\beta ^2)v''v-(v'')^2
\biggr)\cr
&~~~~~~~~~~~~~~~~~~~~~ +\bigl(vv'''-4\beta
vv''-2\beta (1+\beta ^2)v^2
-v'v''+2\beta (v')^2\bigr )\bigr|_{x=\xi,t}~.\cr 
}
\EQ{e17}
$$
We write $B_\xi(t)$ for the boundary term obtained above:
$$
B_\xi(t)\,=\,
\bigl(vv'''-4\beta
vv''-2\beta (1+\beta ^2)v^2
-v'v''+2\beta (v')^2\bigr )\bigr|_{x=\xi,t}~.
$$
Finally, we rewrite \equ{e17} by completing a square:
$$
\eqalign{
\HALF \partial _t \J(t)\,&=\,
\int_\xi^\infty \kd{x} 
\biggl(\bigl(\epsilon ^2-(1+\beta ^2)^2+(1+3\beta ^2)^2\bigr)v^2 - e^{-2\beta (x-\xi
)}v^4 
\cr
&~~~~~~~~~~~~~~~~~~~~~-\bigl(v''+(1+3\beta ^2)v\bigr)^2\biggr) +B_\xi(t)~.\cr 
}
\EQ{e18}
$$ 
Note that \equ{e18} leads immediately to a differential
inequality:
$$
\HALF \partial _t \J(t)\,\le\,G(\beta )\J(t)+B_\xi(t)~,
\EQ{ineq}
$$   
with
$$
G(\beta )\,=\,\epsilon ^2-(1+\beta ^2)^2+(1+3\beta ^2)^2\,=\,\epsilon
^2+4\beta ^2+8\beta ^4~.
\EQ{e22}
$$
This is the origin of the polynomial in \equ{ccrit}. 
We bound first the boundary term.
\CLAIM{Lemma}{boundary}There is a $\C{boundary}$ such that for all $u_0\in\BB$,
all $\xi$, and all $t>0$ one has
$$
B_\xi(t)\,\le\,\C{boundary}~.
\EQ{bb} 
$$

\PROOF Recall that $v_\xi(x,t)=e^{\beta (x- \xi)}u(x,t)$. Using
elementary calculus, we find
$$
\partial _x^j v_\xi(x,t)\,=\,\sum_{k=0}^j \left ({j\atop k}\right )
\beta ^j e^{\beta (x-\xi)}\partial _x^{j-k}u(x,t)~.
$$
Therefore, 
$$
\partial _x^j v_\xi(\xi,t)\,=\,\sum_{k=0}^j \left ({j\atop k}\right )
\beta ^j \partial _\xi^{j-k}u(x,t)|_{x=\xi}~,
$$
and the assertion follows because $u\in\BB$.

Using \clm{boundary}, we conclude from \equ{ineq} that
$$
 \partial _t \J(t)\,\le\,2G(\beta )\J(t)+2\C{boundary}~.
$$
Solving the differential inequality from $t$ to $t'$, we obtain for $t'>t$,
$$
\J(t')\,\le\,e^{2G(\beta )(t'-t)} \J(t) + 2{e^{2G(\beta
)(t'-t)}-1\over 2G(\beta )}\C{boundary}~.
\EQ{bbb} 
$$ 
We need this inequality in a slightly different form.
Note that for ${\xi'} >\xi$, one has
$$
\eqalign{
J_{\xi'} (t)\,=\,\int_{\xi'} ^\infty \d{x} u^2(x,t)e^{2\beta (x-{\xi'} )}\,&=\,
e^{-2\beta ({\xi'} -\xi )}\int_{\xi'} ^\infty \d{x} e^{2\beta
(x-\xi)}u^2(x,t)\cr
\,&\le\,
e^{-2\beta ({\xi'} -\xi )}\int_\xi ^\infty \d{x} e^{2\beta
(x-\xi)}u^2(x,t)\cr
\,&=\,e^{-2\beta ({\xi'}-\xi)}J_\xi(t)~.\cr
}
\EQ{good0} 
$$
Combining this with \equ{bbb} we get for ${\xi'} >\xi$ and ${t'}>t$:
$$
J_{\xi'} ({t'})\,\le\,e^{-2\beta ({\xi'} -\xi )}
\bigl(e^{2G(\beta )({t'}-t)} \J(t) + {e^{2G(\beta
)({t'}-t)}-1\over G(\beta )}\C{boundary}\bigr )~.
\EQ{good} 
$$
To complete the proof of \clm{main}, it suffices to set $\xi'=c\tau$,
$t'=\tau$, $\xi=0$ and $t=0$ in \equ{good}. Then we get
$$
J_{c\tau}(\tau )\,\le\,
e^{2(G(\beta )-\beta c)\tau} \bigl(J_0(0)+{\C{boundary}\over G(\beta
)}\bigr)~.
\EQ{fff} 
$$ 
Clearly, if $c>G(\beta )/\beta $, then $J_{c\tau }(\tau )\to 0$ as
$\tau \to\infty $. Thus, if $J_0(0)<\infty $, and the assertion of
\clm{main} follows.\QED
\REMARK One can do a little better than \equ{fff}. Namely, consider
the case where $c=G(\beta )/\beta $, that is, the case of a critical
speed. Then one finds from \equ{good} that
$$  
J_{c\tau+\lambda}(\tau )\,\le\,e^{-2\beta
\lambda}\bigl(J_0(0)+{\C{boundary}\over G(\beta 
)}\bigr)~,
$$
and in particular $\lim_{\lambda \to\infty} J_{c\tau +\lambda }(\tau )=0$, if
$J_0(0)$ is finite. This means that in the frame moving with exactly
the critical speed, no amplitude ``leaks'' far ahead in that
frame in ${\rm L}^2(e^{2\beta x}\d{x})$.
One can compare this with the results of Bramson [B] who showed
(for positive solutions of the Ginzburg-Landau equation) that such a
leakage is only possible if the initial data decay like $e^{-x}
x^\alpha $ with $\alpha > 1$. In that case, he gets positive
amplitudes at $ct+(\alpha -1)\log t$. Note that the condition
$J_0(0)<\infty $ can only hold for $\alpha < -\HALF$, and then the
correction term will push the amplitude behind the position of
$ct$. Thus, in the case of the Ginzburg-Landau equations the two results
are consistent.

\SECT{nonlin}{An example of a non-linear velocity bound}Consider the
semi-linear parabolic equation 
$$
\partial_{t}u=P(\partial_{x})u+f(u)~,
\EQ{GLL} 
$$
where $P$ is a real polynomial, $\Re P(ik)$ diverges to $-\infty $ as
$|k|\to\infty $ and $\Im(ik)$ is a polynomial of lower
order.\footnote{${}^*$}{The complex Ginzburg-Landau equation is
somewhat more complicated because in that case $P$ is a $2\times2$ matrix
polynomial. But it is covered by our methods.}
We also assume that $f$ is a $\CC^2$
function for which
$f(0)=0$, and $f'(0)=0$. This implies that $u=0$ is an unstable
fixed point of \equ{GLL} . We also assume that
$$
\limsup_{|u|\to\infty}{f(u)\over u}\,<\,0~.
$$
This assumption ensures global existence and regularity of the
semiflow 
(see [CE]).
(If $\vec u$ is
vector valued we impose  $\limsup_{\|\vec u\|\to\infty} \vec u\cdot\vec
f(u)/\|\vec u\|^{2}<0$.) 

Define
$$
\sigma\,=\,\sup_{u}{f(u)\over u}~.  
$$
This is a finite positive quantity from the above assumptions (if $\vec u$ is
vector valued we define it as the sup of $\vec u\cdot\vec f(u)/\|\vec
u\|^{2}$
.)
Note that one can have $\sigma>f'(0)$, and if this happens Aronson and
Weinberger [AW] showed that the minimal speed is bounded above by
$\sqrt{4\sigma}$, when $P(ik)=-k^2$. In this section we show that the
same result can be recovered for this, and many other equations using
the methods of \sec{speed}, again without any recourse to the maximum
principle. 

In this case,
Eq.\equ{e22}  becomes
$$
G(\beta)\,=\,Q(\beta)+\sigma~,
$$ 
where $Q$ is given by
$$
Q(\beta)\,=\,\sup_{k^*_\beta }\Re P(-\beta+ik^{*}_{\beta})
$$
where the $k^*_\beta $ are the solutions of
$$
\left .{{\rm d}\Re P(-\beta+ik)\over \d{k}}\right| _{k=k^*_\beta }\,=\,0~.
$$
The remainder of the proof is the same, except that in \equ{e18} the
term
$-\exp(-2\beta (x-\xi))v^4$ is replaced by  
$$
e^{\beta
(x-\xi)}vf\bigl(e^{-\beta
(x-\xi)}v\bigr)\le \sigma v^2~$$
After this modification the proof proceeds as before.

\SECTIONNONR{Appendix: The determination of the critical
speed}\def\actualnumber{A}Let $P$ 
be a real polynomial 
for which $\Re P(ik)$ diverges to $-\infty $ as
$|k|\to\infty $ and $\Im P(ik)$ is of lower order. In the case of SH, we have
$P(z)=\epsilon ^2-(1+z^2)^2$.
For $\beta>0$ we consider $P(-\beta+ik)$, take the real part and look
for an extremum in $k$. In other words, 
we solve
$$
{{\rm d}\Re P(-\beta+ik)\over \d{k}}\,=\,0~,
$$
in the unknown $k$.
Since $P$ is analytic, one can write this as
$$
0\,=\,\Im\left(\left .{{\rm d} P(z)\over\d{z} }\right |_{z=-\beta
+ik}\right)~. 
\EQ{imcond} 
$$
For each $\beta $ we find solutions $k^*_{\beta}$. 
The 
velocity 
$c^*_\beta$ is related to the
critical value of $P$  in \equ{imcond} by
$$
c^*_\beta \,=\,\sup_{k^*_\beta}\Re P(-\beta+ik^*_{\beta})/\beta~.
\EQ{cstar}
$$
Then, the minimal speed
is
$$
c_*\,=\,\inf_{\beta \in(0,\infty] } c^*_\beta~,
$$
which is determined by \equ{ccc}.
To simplify the discussion, we will assume from now on that for all
$k_\beta ^*$ one obtains the same critical value. This is the case for
the Ginzburg-Landau and Swift-Hohenberg equations.

Note that there is at least one $\beta _*$ solving
$$
\partial _\beta \bigl(\Re P(-\beta+ik^*_{\beta})/\beta\bigr
)\bigr|_{\beta =\beta_*}=0~,
\EQ{ccc} 
$$
for which $c_*=c^*_{\beta_*}$.

In the approach of [BBDKL]
the authors consider $\omega_0(k)=-P(ik)$. They determine
$\bar k(c)\in\complex$ by 
$$
({\rm d}\omega_0/{\rm d}\bar k)|_{k=\bar k(c)}\,=\,i c~,
\EQ{bjl1} 
$$
and then $c_*\in\real$ by
the condition
$$
\Re\bigl( \omega(\bar k(c_*)) -i \bar k(c_*) c_*\bigr ) \,=\,0~.
\EQ{bjl2} 
$$

To compare the two approaches,
note that $P(-\beta+ik)=-\omega_0(k+i\beta)$.
Clearly the equations \equ{cstar} and \equ{bjl2} are equivalent.
To see that \equ{imcond} and \equ{bjl1} say the same thing, note that  
since $c$ is real one has
$$
\Im(P'(z))\,=\,\Re(-\omega_0'(-iz))\,=\,\Re(-\omega_0'(-iz)+ic)~.
\EQ{equal} 
$$
In particular, if $\omega_0$ is an even function, the relation
$\Re(\omega_0'(\bar k)-ic)=0$ is equivalent to requiring
$\omega_0'(\bar k)=ic$, which is \equ{bjl1}. 
Using \equ{equal}, we conclude 
that the solution $\bar k$ of $\omega_0'(\bar k)=ic$ of [BBDKL] is the
same as $-i$ times the solution $z$ of $\Im(P'(z))=0$, which is \equ{imcond}. 
Therefore $\bar k =k^*_{\beta_*} +i\beta_*$. 
Finally, to find $c^*_{\beta_*} $ one can solve
$$
\displaylines{
0\,=\,\Re(P(-{\beta_*}+ik^*_{{\beta_*}})-{\beta_*} c^*_{\beta_*}
)\,=\,\Re(-\omega_0(k^*_{\beta_*} +i{\beta_*})-ic^*_{\beta_*} (k^*_{\beta_*} +i{\beta_*}))\cr\,=\,
\Re(-\omega_0(\bar k )-ic^*_{\beta_*} \bar k )~.
}
$$
\REMARK The same kind of calculation can be done for multi-component
problems (such as reaction diffusion), where $P$ would be a matrix.
\LIKEREMARK{The example of the SH equation}In this case
$$
P(z)\,=\,\epsilon ^2 -(1+z^2)^2~,
$$
and so 
$$
\omega_0(k)\,=\,-P(ik)\,=\,-\epsilon ^2 +(1-k^2)^2~.
$$
In [BBDKL], it is found that
$$
\eqalign{
\bar k \,&=\, \bar k_1 + i\bar k_2~,\cr
\bar k_2 \,&=\, {\sqrt{\sqrt{1+6\epsilon ^2}-1}\over 12}\,=\,{\epsilon
^2\over 4}+\dots~,\cr
\bar k_1\,&=\,1+3\bar k_2^2~,\cr
c_*\,&=\, 8 \bar k_2 (1+4 \bar k_2^2)\,=\,4\epsilon +\dots~. 
}
\EQ{BJL} 
$$
In our formulation, we find 
$$
P(-\beta +ik)\,=\,\epsilon ^2 -\bigl(1+(ik-\beta )^2\bigr )^2~.
$$
The real part of the derivative w.r.t.~$k$ yields
$$
{{\rm d}\Re P(-\beta+ik)\over \d{k}}\,=\,4k -4k^3 +12 k\beta ^2~.
$$
The solutions of ${{\rm d}\Re P(-\beta+ik)/\d{k}}=0$ are
$k^*_\beta =\pm\sqrt{1+3\beta ^2}$ (and $k^*_\beta =0$ which leads to
less stringent bounds). Substituting back into $\Re P$, we
get
$$
\Re P(-\beta +ik^*_\beta )\,=\,\epsilon ^2+4\beta ^2+8\beta
^4~,
$$
which is what we announced in \equ{newcbeta} and got  as a result of
integration by parts in Eqs.\equ{e18}--\equ{e22}.  
Solving now
$$
\Re P(-\beta +ik^*_\beta )-c\beta \,=\,0~,
$$
for $c=c^*_\beta  $ leads to $c^*_\beta  =(\epsilon ^2+4\beta ^2+8\beta
^4)/\beta $. To find the absolutely minimal speed, we find that $\beta
$ for which $c^*_\beta  $ is extremal, that is $\partial _\beta
c^*_\beta =0$. The only positive solution is
$$
\beta_*\,=\,{\sqrt{3\sqrt{1+6\epsilon ^2}-3}\over 6}~,
$$
and hence,
$$
c^*_{\beta _*}\,=\,{
4\bigl(\sqrt{1+6\epsilon ^2}-1+6\epsilon ^2\bigr) \over
3\sqrt{3\sqrt{1+6\epsilon ^2}-3}}~. 
$$
This quantity is the same as $\inf_{\beta\in\real}  c_\beta $ where $c_\beta $
is given by \equ{ccrit}.
\LIKEREMARK{Acknowledgments}We thank W. van Saarloos for some help
with the references, and M. Hairer and G. van Baalen for a critical
reading of the manuscript.
This work was partially supported by the
Fonds National Suisse.
\SECTIONNONR{References}

\eightpoint\frenchspacing\raggedright
\widestlabel{[BBDKL]}
\ref
\no{AW}
  \by{Aronson, D. and H.F. Weinberger}
  \paper{Multidimensional nonlinear diffusion arising in population genetics}
  \jour{Adv. Math.}
  \vol{30}
  \pages{33--76}
  \yr{1978}
\endref
\ref 
\no{BBDKL}
 \by{Ben-Jacob, E., H. Brand, G. Dee, L. Kramers, and J.S. Langer}
 \paper{Pattern propagation in non-linear dissipative systems}
 \jour{Physica}
 \vol{14D}
 \pages{348--364}
 \yr{1985}
\endref
\ref
\no{B}
\by{Bramson, M.} 
\paper{Convergence of solutions of the Kolmogorov equation to travelling
waves.}
\jour{Mem. Amer. Math. Soc.}
\vol{44} 
\yr{1985}
\pages{No. 285, iv+190pp}
\endref 
\ref
\no{CE}
  \by{Collet, P. and J.-P. Eckmann}
  \book{Instabilities and Fronts in Extended Systems}
  \publisher{Princeton University Press}
  \yr{1990}
\endref
\ref
\no{DL}
\by{Dee, G. and J. Langer}
\paper{Propagating pattern selection}
\jour{Phys. Rev. Lett.}
\vol{50}
\pages{383--386}
\yr{1983}
\endref
\ref
\no{LMK}
 \by{Langer, J. and H. M\"uller-Krumbhaar}
     \paper{Mode-selection in a dendrite-like nonlinear system}
\jour{Phys. Rev.} 
\vol{A 27}
\pages{499--514}
\yr{1983}
\endref
\ref
\no{vS}
 \by{van Saarloos, W.}
     \paper{Front propagation into unstable states: Marginal stability
     as a dynamical mechanism for velocity 
         selection}
\jour{Phys. Rev.} 
\vol{A 37}
\pages{211--229}
\yr{1988}
\endref
\bye